\documentclass[11pt,a4paper]{article}
\usepackage[dvips]{graphicx}
\usepackage{subfigure}
\usepackage{latexsym}
\input epsf
\begin{document}

\title{The commensurate-disordered phase transition in 2D classical ATNNI model studied by DMRG}
\author{A.~Gendiar and A.~\v{S}urda\\
Institute of Physics, Slovak Academy of Sciences, D\'{u}bravsk\'{a} cesta 9,\\
SK-842 28 Bratislava, Slovakia} 

\maketitle{}

\begin{abstract}
The classical two-dimensional anisotropic triangular nearest-neighbor Ising (ATNNI)
model is studied by the density matrix renormalization group (DMRG) technique
when periodic boundary conditions are imposed. Applying the finite-size scaling 
to the DMRG results a commensurate-disordered (C-D) phase transition line as well as 
temperature and magnetic critical exponents are calculated.
We conclude that the (C-D) phase transition in the ATNNI model belongs to the
same universality class as the ordered-disordered phase transition of the Ising model.
\end{abstract}

\renewcommand{\theequation}{\arabic{equation}}
\setcounter{equation}{0}

Analysis of semi-finite systems of small size in one or more directions
 has been used as a powerful tool in
extracting of critical properties of  two-dimensional classical models 
and corresponding one-dimensional quantum models. Although finite  or
1D systems themselves do not
display any critical behavior, it is, however, possible to extract critical parameter 
values as well as critical exponents. Temperature, ordering magnetic field, and finite-size
deviations from criticality are all described by the same set of the critical exponents 
\cite{Nig}. This paper is focused on infinite strips of finite width  where 
the relevant numerical data are 
obtained from the transfer matrix methods, in particular, the Density Matrix Renormalization 
Group (DMRG) method.

In 1992 the DMRG technique has been invented by S.~R.~White \cite{Whi} in real space for 
one--dimensional (1D) quantum spin Hamiltonians. Three years later T. Nishino \cite{Nis} 
applied this numerical technique to classical spin 2D models that is based on the 
renormalization group transformation for the transfer matrix for the open boundary 
conditions. 
DMRG treatment of
 2D classical systems  exceeds the classical Monte Carlo approach
in accuracy, speed, and size of the systems \cite{Kar}.

Recently, we have modified the DMRG method for the 2D classical models,
imposing periodic
boundary conditions (PBC) on strip boundaries, and found a relation that 
helped to determine an optimal
strip width  $L^{\rm opt}$ in order to obtain  correct  values of critical
temperature and 
exponents \cite{Ag1} using the finite-size scaling (FSS). We have obtained results
of very high accuracy exceeding  the DMRG method with standard open
boundary conditions. Our method does not require any extrapolation analysis of the data.

The use of DMRG for 2D classical models may follow one of two different approaches:

(i)  DMRG method is applied to strips of finite width and from
two largest transfer-matrix eigenvalues  or the free energy estimated
with high precision, the critical properties of the system are calculated 
by the FSS analysis (here, we use this approach).

(ii) The strip width is enlarged until a steady state is reached (in the thermodynamic limit) 
when the output from the DMRG does not depend on the lattice size. Then, the DMRG
yields properties of the 2D infinite system with spontaneously broken symmetry
and  mean-field-like behavior close to the criticality. This approach was
used recently to study the high-field part of the ATNNI-model phase diagram
\cite{Ag2},
where  approach (i) ran into convergence problems. We were able to show that the phase
transition between the commensurate phase and the disordered phase proceeds
via a narrow strip of an incommensurate phase. This approach gives also
accurately the low-field  part of the  phase diagram, but it is not
convenient for determination of the critical properties of the system by FSS.
In distinction to the finite-width approach (i), it
explicitly undergoes the phase transition, but its critical behavior is
mean-field-like and the speed of calculation suffers from critical
slowing-down at the phase transition line. Therefore, we use here approach (i) to
find the low-field critical behavior of the ATNNI model.

The FSS approach should give the correct critical properties of the system in the limit 
of infinite strip width. Nevertheless, it was shown
in \cite{Ag1} that in approximate DMRG treatment  for given size of the transfer matrix 
(limited by computer capacity), it is not useful to enlarge the  strip width 
to too large  values, because here the the DMRG results  do not satisfy the scaling laws 
assumed by the FSS. Thus, an optimal width, up to which the results
systematically improve, must exist. It was also shown that the estimation of critical properties 
of the Ising and Potts models by DMRG with the periodic boundary conditions are much better
than those with the open ones, despite the latter yields better results for the
finite-width strips \cite{Ag1}.

Below the optimal strip width $L^{\rm opt}$ the ratio   
\begin{equation}
R\equiv\frac{\frac{\partial}{\partial L}T_{\rm C}^*(L)}{\frac{{\partial}^2}{\partial L^2}
T_{\rm C}^*(L)}
\end{equation}
is almost linear function of $L$ while above it, it is not.  
($L$ in (1) is the width of the strip  and $T_{\rm C}^*(L)$ is the 
critical temperature for given $L$.) 

The deviation of $R$ from linearity  above the optimal strip width is  very fast 
and the ratio $R$ becomes 
 zero or infinity within enlargement of the strip by one lattice constant. 
Thus, if $R=0$ or $R\to\infty$ (i.e. if the numerator
or the denominator tends to zero or changes its sign), we accept that $L$ 
as the strip width for further calculations and call 
it the optimal width $L^{\rm opt}$ of the strip. 
The critical temperature for the optimal width 
$T_{\rm C}^*(L^{\rm opt})$ is taken as the best approximation of the
critical temperature of the 2D system  studied, and at this temperature the
critical exponents of the system are calculated. The  values of the
critical exponents are sensitive
to $T_{\rm C}^*$ and must be determined with the due care.

In the FSS approach, the critical exponents are derived from the scaling
behavior of the correlation length and free energy at critical point, where they
depend on strip width $L$ in the following  way \cite{Nig}:
\begin{equation}
K_L^{\rm h}\sim L^{2y_{\rm h}^{(\beta)}}
\qquad
K_L^{\rm T}\sim L^{y_{\rm T}^{(\nu)}}
\qquad
c_{\rm L}\sim L^{2y_{\rm T}^{(\alpha)}-d}
\label{e1}
\end{equation}
where $K_L^{\rm T}$  and $K_L^{\rm h}$ are the derivatives of inverse
correlation length $K$ with respect to temperature $T$  and second
derivative with respect to ordering (magnetic)
field $h$, respectively, and $c_{\rm L} $  is the specific heat, i.e. the
second derivative of the free energy with respect to temperature. 
The two temperature exponents $y_{\rm T}^{(\alpha)}$ and  $y_{\rm T}^{(\nu)}$
should be equal to each other.
The  exponents 
$y_{\rm T}$ and $y_{\rm h}$ determine the critical behavior of all 
statistical  quantities
 characterizing the system. The critical exponents of specific heat,
magnetization and correlation length can be calculated from  
$y_{\rm T}$ and $y_{\rm h}$ as follows: 
$\alpha=2-\frac{2}{y_{\rm T}}$, $\beta=\frac{2-y_{\rm h}}{y_{\rm T}}$,
$\nu=y_{\rm T}^{-1}$.
Other critical exponents can be obtained from the 
scaling equations $y_{\rm h}=\beta+\gamma$, $\gamma=\beta(\delta-1)$, and 
$\eta=2-\gamma y_{\rm T}$ \cite{Xer}.

Further, we demonstrate the capabilities of our approach to find 
 the critical properties of 2D spin lattice model on Ising model with
different symmetries of the lattice, where critical temperatures and
critical indices are known from exact solutions, and ATNNI model where the
phase diagram is generally unknown and the critical indices are
predicted from symmetry considerations. 

\begin{figure}[!ht]
\centering
\subfigure[\hfill]{\scalebox{0.5}{\includegraphics{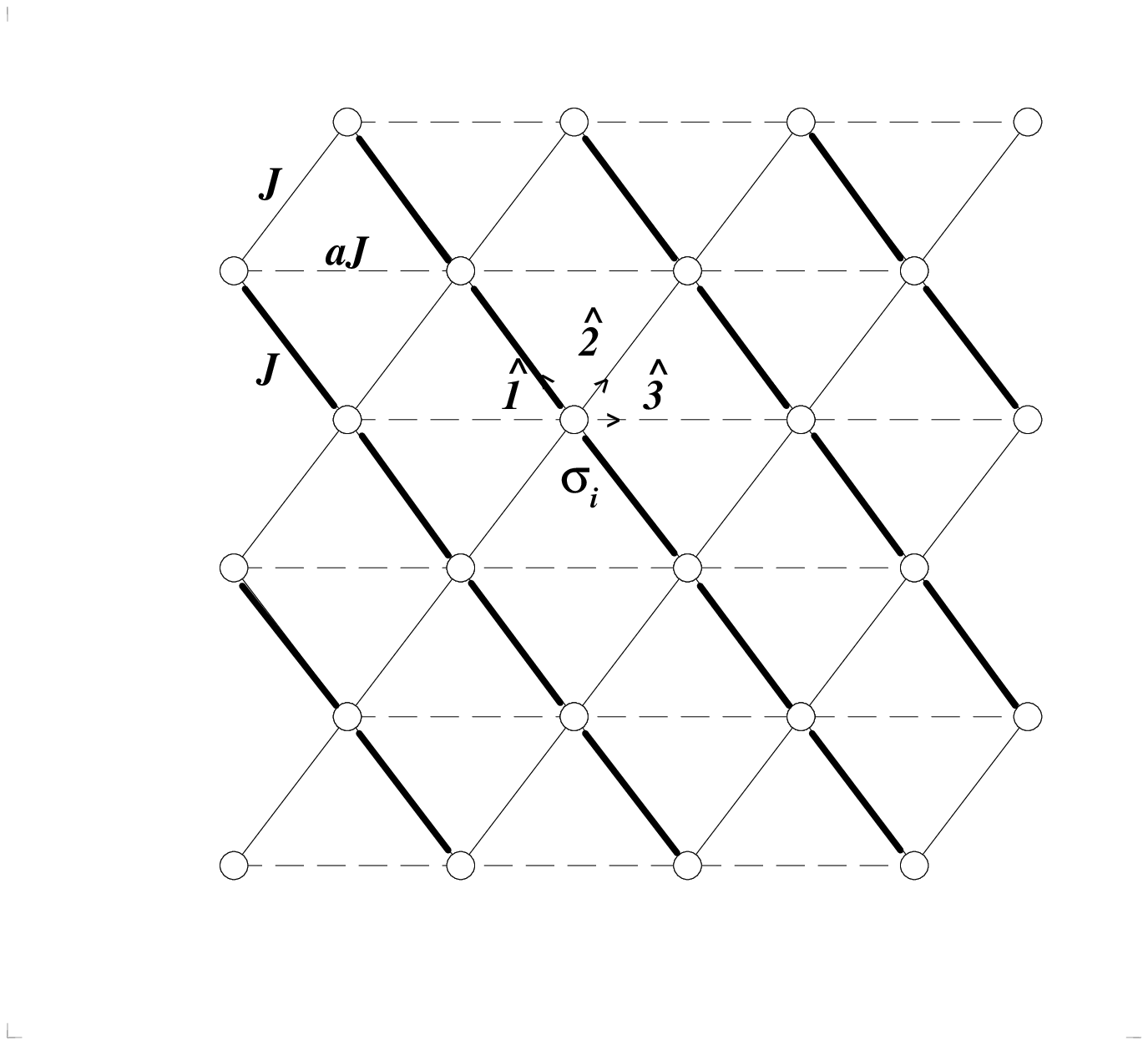}}}
\hfill
\subfigure[\hfill]{\scalebox{0.5}{\includegraphics{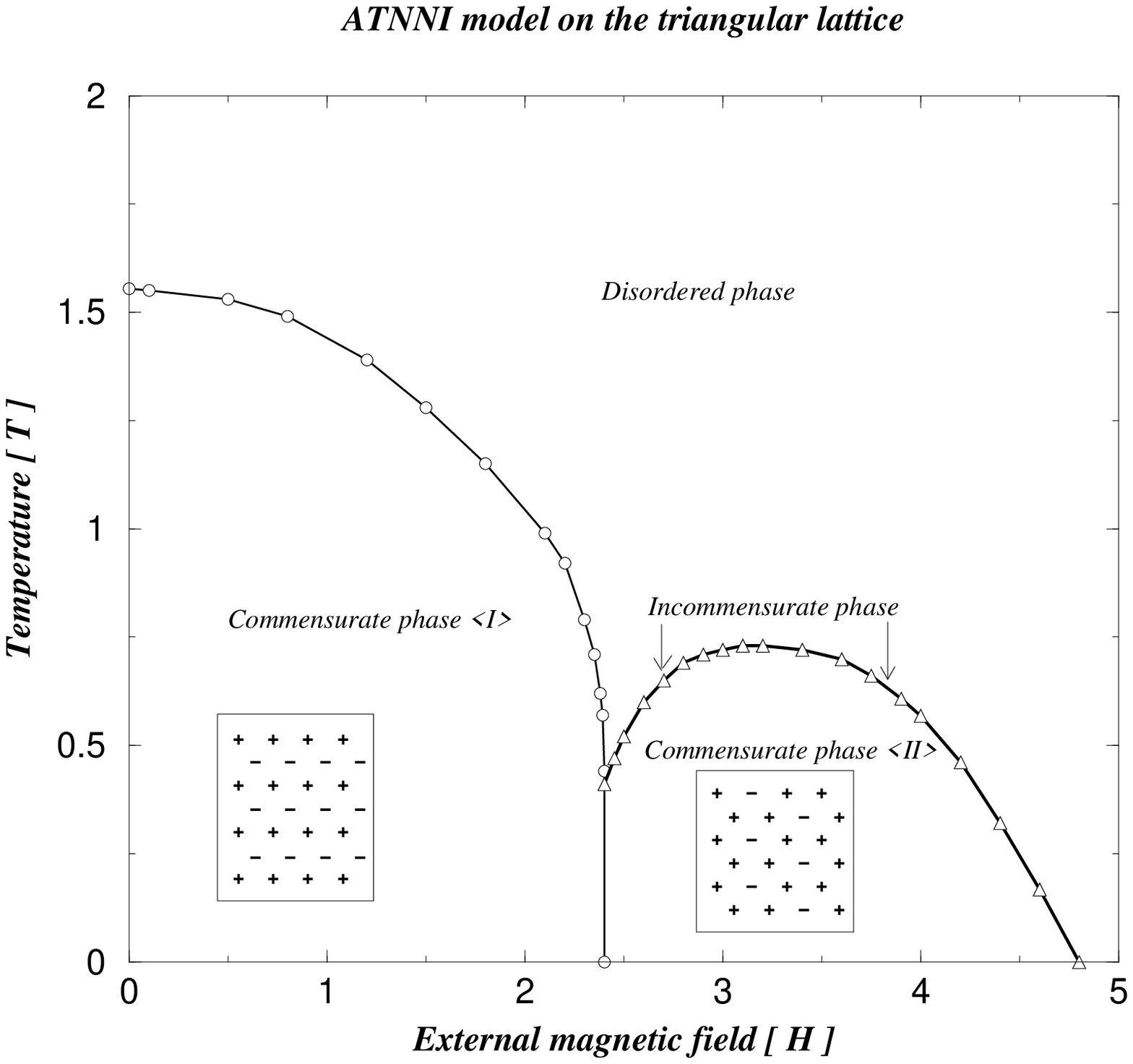}}}
\caption{\it (a) The triangular lattice of the ATNNI model. (b) The phase diagram
of the ATNNI model for a=0.4 obtained by DMRG \cite{Ag2}.}
\label{f1}
\end{figure}

The 2D classical anisotropic triangular nearest-neighbor Ising (ATNNI) model
is given by the Hamiltonian
\begin{equation}
{\cal H}=\sum\limits_{i}-J\left(\sum\limits_{\hat\delta=\hat 1,\hat 2} \sigma_i
\sigma_{i+\hat\delta}+a \sigma_i \sigma_{i+\hat 3}\right)-H\sum\limits_i \sigma_i
\end{equation}
with the antiferromagnetic coupling $J<0$ and spins $\sigma_i=\pm 1$. 
The numbers $\hat 1$, $\hat 2$, and $\hat 3$ are lattice directions in the ATNNI model. 
The coupling J is multiplied by the parameter $a$ ($0<a<1$)
along the direction $\hat 3$ as depicted in Figure \ref{f1} (a).

This model was studied by Domany and Schaub \cite{Dom} and  in \cite{Ag2},
and it was shown that its phase diagram, as a plot of temperature $T$ 
and external magnetic field $H$ (for $a=0.4$), exhibits four different phases: 
two commensurate phases $\langle I\rangle$ and $\langle II\rangle$, a disordered phase, 
and an incommensurate phase, see Figure \ref{f1} (b). Commensurate phase $\langle I\rangle$ 
occurs at magnetic field $H<2.4$. This structure satisfies the Lifshitz condition, and
it is characterized by a one-dimensional representation of the lattice symmetry
group, i.e.  its phase transition is predicted to belong to the Ising universality
class \cite{DSW}. Domany and Schaub tried to confirm this prediction by
numerical calculation of the exponent $y_{\rm T}$, but due to the low-order
approximation it differed  from the expected value  
by more than 10\% and the magnetic exponent was not calculated at all.     

We have calculated critical properties of the ATNNI model at the phase
transition line  between the commensurate  $\langle I\rangle$  and
disordered phase. To illustrate the accuracy of the method, we have
calculated critical properties of the exactly solvable models: ferromagnetic
and antiferromagnetic Ising model on square and triangular lattices at zero
magnetic field as well as the zero-magnetic-field ATNNI model which critical
temperature is given by the equation  
\cite{Bax}
\begin{equation}
\sinh^2\left(\frac{2J}{T_{\rm C}}\right)=\exp\left(-\frac{4aJ}{T_{\rm C}}\right).
\label{Tcrit}
\end{equation}
We have used the FSS analysis of DMRG results with superblock consisting
of 8 Ising spins and 4 multi-spin variable, each acquiring 85 values $(m=85)$. The
computational effort at this approximation is less than for the classical
transfer matrix method of strip width equal to 17 lattice constants. However, the DMRG
enables to treat wider strip (of tens of lattice constants) up to the
optimal width further improving the values of the critical parameters.   

The first, important step of the calculations is determination of the critical
temperature $T^*_{\rm C}$, see Table \ref{tt}, of which the best estimation for given 
$m$ is $T^*_{\rm C}(L^{\rm opt})$ 
calculated from FSS approach \cite{Ag1}. At this temperature the
values of the critical exponents are derived from the scaling laws (2).

\begin{table}[!ht]
\caption {Critical temperatures $T_{\rm C}^*$ obtained from (1) with DMRG compared
with the exact ones $T_{\rm C}^{\rm (exact)}$. The symbols $\Box$ and $\triangle$ describe
square and triangular lattices, respectively.}
\vskip 0.2truecm
\centerline{
\begin{tabular}{|l|c|l|l|}
\hline
{\bf model}  & $H$ & \ \ \quad $T_{\rm C}^*$ & \ \ $T_{\rm C}^{\rm (exact)}$ \\
\hline
$\Box$ Ising & 0.0 & 2.2691851 & 2.2691853  \\
\hline
$\Box$ AF Ising & 0.0 & 2.2691848 & 2.2691853  \\
\hline
$\triangle$ Ising & 0.0 & 3.640955  & 3.640957   \\
\hline
$\triangle$ ATNNI & 0.0 & 1.55352   & 1.55362 \\
\hline
$\triangle$ ATNNI & 0.5 & 1.52867 & unknown \\
\hline
$\triangle$ ATNNI & 1.0 & 1.45135 & unknown \\
\hline
$\triangle$ ATNNI & 1.5 & 1.31105 & unknown \\
\hline
$\triangle$ ATNNI & 2.0 & 1.07009 & unknown \\
\hline
\end{tabular}
\label{tt}
}
\end{table}

\begin{figure}[!ht]
\centering
\subfigure[\hfill]{\scalebox{0.4}{\includegraphics{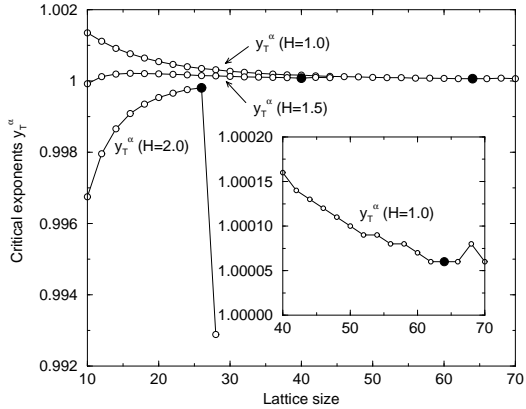}}}
\hfill
\subfigure[\hfill]{\scalebox{0.4}{\includegraphics{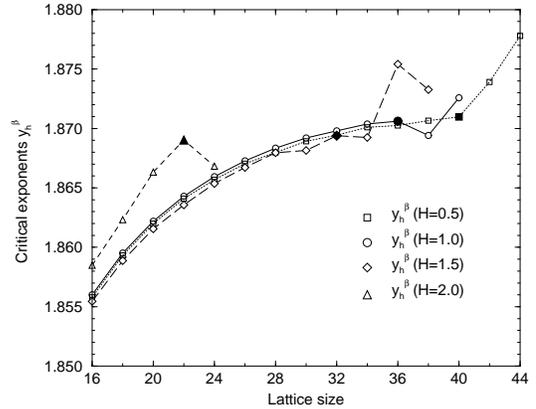}}}
\caption{\it (a) The plot of thermal critical exponents $y_{\rm T}^{(\alpha)}$
for different fields in the ATNNI model. 
(b) The plot of magnetic critical exponents
$y_{\rm h}^{(\beta)}$. The filled symbols denote the accepted
critical exponents satisfying Eq. (1)}
\label{f2}
\end{figure}

As the quantities appearing in (2) are first and second derivatives of the
free energy and correlation length, the effect of approximation starts to
manifest at lower strip width than $L^{\rm opt}$. The criterion determining
strip width at which the value of the critical exponent may be still acceptable, was
taken completely analogous to that for critical temperature, Eq.~(1). The
accepted values of the critical exponents are denoted by filled symbols in Figs \ref{f2}
(a) and (b).

The critical exponent $y_{\rm T}$ is determined more precisely from the free
energy $y_{\rm T}^{(\alpha)}$ than from the correlation length $y_{\rm T}^{(\nu)}$, 
as for the evaluation of the former
only the largest eigenvalue of the superblock matrix is needed in distinction
to the correlation length, to whose calculation the ratio of the largest and
the second largest eigenvalue is necessary. This point is irrelevant for the
models with a symmetric transfer matrix (Ising models in Table~\ref{tt}), but
significant for the  ATNNI model with a non-symmetric transfer matrix \cite{Non}.    
The plot of thermal critical exponent $y_{\rm T}^{(\alpha)}$ vs.
strip width is shown in Figure \ref{f2} (a). For increasing lattice
size they both tend to the Ising value 1. 
The convergence also depends on the magnetic field. It gets worse for
magnetic field close to the multi-critical point $H=2.4$. Here the
reliability of the DMRG breaks down at rather small strip width, as well.
The accepted values depicted by black symbols are listed in Table~\ref{tb}. 

\begin{table}[!ht]
\caption {Critical exponents of various 2D spin models
calculated by the DMRG method with PBC and FSS analysis. 
The exact critical exponents of the Ising models are as follows: $y_{\rm T}=1$ 
and $y_{\rm h}=1.875$.}
\vskip 0.2truecm
\centerline{
\begin{tabular}{|l|c|c|l|l|l|c|l|}
\hline
{\bf model}  & $H$ &
$y_{\rm T}^{(\alpha)}$ & \qquad $y_{\rm T}^{(\nu)}$ & \quad $y_{\rm h}^{(\beta)}$ & 
\qquad $\alpha$ & $\beta$ & \qquad $\nu$ \\
\hline
$\Box$ Ising & 0.0 & 1.0000009 & 0.99999994 & 1.875002
& 0.0000017 & $\frac{1}{8.00012}$ & 1.00000006 \\
\hline
$\Box$ AF Ising & 0.0 & 1.0000009 & 0.99999994 & 1.875126
& 0.0000017 & $\frac{1}{8.00804}$ & 1.00000006 \\
\hline
$\triangle$ Ising & 0.0 & 1.0000014 & 0.99999943 & 1.875030
& 0.0000027 & $\frac{1}{8.00192}$ & 1.00000057 \\
\hline
$\triangle$ ATNNI & 0.0 & 1.0000022 & 0.9947 & 1.87005
& 0.000004 & $\frac{1}{7.70}$ & 1.00527 \\
\hline
$\triangle$ ATNNI & 0.5 & 1.0000280 & 0.9902 & 1.87098
& 0.000056 & $\frac{1}{7.75}$ & 1.00993 \\
\hline
$\triangle$ ATNNI & 1.0 & 1.0000580 & 0.9902 & 1.87062
& 0.000116 & $\frac{1}{7.73}$ & 1.00993 \\
\hline
$\triangle$ ATNNI & 1.5 & 1.0000767 & 0.9911 & 1.86939
& 0.000153 & $\frac{1}{7.66}$ & 1.00893 \\
\hline
$\triangle$ ATNNI & 2.0 & 0.9998366 & 1.0122 & 1.86902
& 0.000327 & $\frac{1}{7.63}$ & 0.98795 \\
\hline
\end{tabular}
\label{tb}
}
\end{table}

The critical exponent $y_{\rm h}^{(\beta)}$  describes the decay of the order
parameter at the phase transition line from the commensurate phase
$\langle I \rangle$ to the disordered phase. The structure  $\langle I \rangle$  
consists of two ferromagnetically ordered  sublattices each with different
magnetization. As the external magnetic field $H$ is  generally non-zero in
ATNNI model, the total magnetization (sum of both sublattice magnetizations)
is non-zero, as well. The difference between the two magnetization is taken
as the order parameter in this case. The small ordering field  $h$  used for
calculation of the derivative $K_L^h$ acquires opposite sign at each of the
two  sublattices. The accuracy of the calculations of the magnetic exponent is
smaller  than that of the thermal exponent in case of exactly solvable
models listed in Table~\ref{tb}. Thus, we can expect a lower accuracy also for ATNNI
model.  All the exponents depicted in Fig. \ref{f2} (b) are below 1.871. Extrapolations
to $L\to \infty$ for $H=0.5$--1.5 give values of $y_{\rm h}^{(\beta)}$  about 1.872,
i.e. $\beta=\frac{1}{7.81}$, which still differs from the Ising value $y_{\rm h}^{(\beta)}=1.875$ 
and corresponding $\beta=\frac{1}{8}$. Note that the value of $y_{\rm h}^{(\beta)}$ is extremely sensitive to the
correct determination of the critical temperature. A very small decrease of
its value would  shift  $y_{\rm h}^{(\beta)}$ to the expected Ising value. At modest
magnetic field, where our calculations are assumed to be more accurate, the
plots of  $y_{\rm h}^{(\beta)}$  for different magnetic field lie on the same curve
what suggests that not only $y_{\rm h}^{(\beta)}$ is a universal quantity
independent of $H$, but the corrections to it for finite $L$ are universal, as
well.

In conclusion, it can be stated that the DMRG method with periodic boundary
conditions  reproduces with a high accuracy the critical properties of
exactly solvable models and confirms the prediction  that the
C-D phase transition for magnetic fields $H=0$--2.4 belongs to the
universality class of the Ising model.    

A.~G. thanks to T.~Nishino for the fruitful correspondence and help.
Most of this work has been done by Compaq Fortran on the HPC Alpha 21264 in Kobe
University in Japan. This work has been supported by Slovak Grant Agency, 
VEGA n. 2/7174/20.



\begin{thebibliography}{99}
\bibitem{Nig} Nightingale P 1982 J. Appl. Phys. {\bf 53} 7927
\bibitem{Whi} White~S~R 1992 Phys. Rev. Lett. {\bf 69} 2863; 1993 Phys. Rev. 
B {\bf 48} 10345
\bibitem{Nis} Nishino T 1995 J. Phys. Soc. Jpn. {\bf 64} 3598
\bibitem{Kar} Hallberg K cond--mat/9910082
\bibitem{Ag1} Gendiar~A and \v{S}urda A submitted to Phys. Rev. B (cond-mat/0004024)
\bibitem{Ag2} Gendiar~A and \v{S}urda A June 2000 Phys. Rev. B (in press), (cond-mat/9912131)
\bibitem{Xer} Balescu R 1978 {\it Equilibrium and nonequilibrium statistical mechanics} (Moscow: Mir)
\bibitem{Dom} Domany~E and Schaub~B 1983 Phys. Rev. B {\bf 29} 4095
\bibitem{DSW} Domany~E, Schick~M and Walker~J~S 1977 Phys. Rev. Lett. {\bf 38} 1148;
Domany~E, Schick~M, Walker~J~S and Griffiths~R~B 1983 Phys. Rev. B {\bf 18} 2209
\bibitem{Bax} Baxter~R~J 1982 {\it Exactly Solved Models in Statistical Physics} 
(London: Academic Press)
\bibitem{Non} Nishino~T and Shibata~N 1999 J. Phys. Soc. Jpn. {\bf 68} 3501;
Shibata~N 1997 J. Phys. Soc. Jpn. {\bf 66} 2221; Wang~X and Xiang~T 1997 Phys.Rev. B 
{\bf 56} 5061; Burssil~R~J, Xiang~T and Gehring~G~A 1996 J. Phys. Cond.
 Mat. {\bf 8} L583; Maisinger~K and Schollw\"{o}ck~U 1998 Phys. Rev. Lett. {\bf 81} 445

\end{thebibliography}
\end{document}